\documentclass[conference,a4paper]{IEEEtran}
\usepackage{cite}
\usepackage{makecell}
\usepackage[utf8x]{inputenc}
\usepackage{graphicx}
\usepackage{subfigure}
\usepackage{array}
\usepackage{booktabs}
\usepackage{url}
\usepackage[misc]{ifsym}
\usepackage[dvips]{color}
\usepackage{mathrsfs}
\usepackage[ruled,linesnumbered]{algorithm2e}
\usepackage{amsmath}
\usepackage{amsfonts,amssymb}
\usepackage{overpic}
\usepackage{multirow}

\makeatletter
\renewcommand{\@pnumwidth}{1.75em}
\renewcommand{\@tocrmarg}{2.75em}
\makeatother

\begin{document}
\title{Phase-based Variant Maximum Likelihood Positioning for Passive UHF-RFID Tags}
\author{\IEEEauthorblockN{
Chenglong~Li\IEEEauthorrefmark{1}$^\star$,   
Emmeric~Tanghe\IEEEauthorrefmark{1},   
David~Plets\IEEEauthorrefmark{1},    
Pieter~Suanet\IEEEauthorrefmark{2},      
Nico~Podevijn\IEEEauthorrefmark{1},\\   
Jeroen~Hoebeke\IEEEauthorrefmark{3},   
Eli~De Poorter\IEEEauthorrefmark{3},    
Luc~Martens\IEEEauthorrefmark{1},    
Wout~Joseph\IEEEauthorrefmark{1}      
}                                     

\IEEEauthorblockA{\IEEEauthorrefmark{1}
WAVES group, Department of Information Technology, Ghent University-IMEC, 9052 Ghent, Belgium}
\IEEEauthorblockA{\IEEEauthorrefmark{2}
Aucxis cvba, 9190 Stekene, Belgium}
\IEEEauthorblockA{\IEEEauthorrefmark{3}
IDLab group, Department of Information Technology, Ghent University-IMEC, 9052 Ghent, Belgium} 
 \IEEEauthorblockA{$\star$ Email: \emph{chenglong.li@ugent.be} }
}

\maketitle
\begin{abstract}
Radio frequency identification (RFID) technology brings tremendous advancement in Internet-of-Things, especially in supply chain and smart inventory management. Phase-based passive ultra high frequency RFID tag localization has attracted great interest, due to its insensitivity to the propagation environment and tagged object properties compared with the signal strength based method. In this paper, a phase-based maximum-likelihood tag positioning estimation is proposed. To mitigate the phase uncertainty, the likelihood function is reconstructed through trigonometric transformation. Weights are constructed to reduce the impact of unexpected interference and to augment the positioning performance. The experiment results show that the proposed algorithms realize fine-grained tag localization, which achieve centimeter-level lateral accuracy, and less than 15-centimeters vertical accuracy along the altitude of the racks.
\end{abstract}

\IEEEpeerreviewmaketitle

\section{Introduction}
Advancements in Internet-of-Things (IoT) technology and integrated circuit hardware have greatly stimulated the growth of industrial IoT. Inventory represents a significant portion of assets in a business, so accurate and reliable data are essential for an efficient and effective business operation. The increasing success of supply chain business requires flexible and continuous inventory management in smart factories, supermarkets, etc. Nowadays, most warehouse systems have adopted automatic identification technology such as radio frequency identification (RFID) tags for automated inventory control. The technical advances in passive RFID-based localization have resulted in enhanced performance in fast, accurate and convenient inventory management.
\par
However, modulated backscatter ultra-high frequency (UHF) RFID is a short range and narrow bandwidth connection of only tens of Megahertz, so time (difference) of arrival is non-realistic. The received signal strength indicator (RSSI)-based technique was considered in \cite{Zhang2014,Khan2009} for its low complexity. However, RSSI is severely affected by the propagation environment, the absorption and scattering, as well as antenna effects such as impedance mismatch and polarization mismatch. This can reduce the power observed at the reader receiver. Multipath propagation and undesired signals in the environment can combine with the primary backscatter \cite{Khan2009}, thereby increasing or decreasing the received signal power at the reader receiver. To realize fine-grained localization with narrowband RFID, the phase-based positioning methods in time, frequency, and space domain were proposed for the first time in \cite{Nikitin2010}. In \cite{Buffi2015,Buffi2017}, stemming from the concept of synthetic aperture radar (SAR), a phase-based localization technique for UHF-RFID tags moving on a conveyor belt was investigated. \cite{LiuTianCi} discussed the possibility of anchor-free phase-based positioning for RFID tags for static applications based on hyperbolic positioning. The interval constraint between the adjacent antennas (less than half a wavelength) was added to solve the phase ambiguity. However, multipath propagation always exists, especially in metallic warehouses \cite{Tanghe2008}. This may contaminate the measured phase and affect the localization accuracy severely. In \cite{Zhao2017,Tagoram}, a probability based weight was constructed, which assigned different ratios for different sampling positions to relieve the impact of multipath propagation. \cite{Shangguan2016} utilized multi-frequency based holography to localize the tags and suppress multipath. Specifically, a weighted function is built based on the normalized information entropy of phase differences between two consecutive channels. 
\par
In this paper, we propose a phase-based passive tag positioning method based on maximum likelihood estimation (MLE). The likelihood function and the weight are reconstructed to mitigate the phase uncertainty and improve the positioning performance. The remainder of this paper is organized as follows. Section II presents the signal model, the performance of the measured phase for different orientations, and the statistical distribution. In Section III, a phase-based maximum likelihood positioning method is derived, and the likelihood function reconstruction is conducted to improve the performance of localization. The experiment and results are shown in Section IV. Finally, Section V concludes this paper.

\section{UHF-RFID Phase Investigation}
\subsection{Signal Model}
Considering a narrowband RFID reader with transceiver co-located, and a symmetric channel for up-/down-link, as shown in Fig. 1, the transmitted signal at time $t$ is given as
\begin{equation}\label{1}
{s_{Tx}}(t)={a_{Tx}(t)}\cos \left(2\pi {f_0}t+{\varphi _{Tx}}\right),
\end{equation}where $a_{Tx}$ denotes the amplitude, $f$ the carrier frequency, $\varphi _{Tx}$ the initial phase shift (caused by the transmitter's circuits and antenna) at transmitter. After the round-trip propagation, the backscattered signal can be given by
\begin{equation}\label{2}
{s_{Rx}}\!(t)\!=\!{a_{Rx}(t)}\cos \left(2\pi f(t\!-2\tau_0)\!+\!{\varphi_{Tx,Rx}}\right){s_{ID}}\!({t\!-\!\tau_0 }),
\end{equation}where ${a_{Rx}}$ is the received amplitude, ${\tau_0}=\frac{d}{c} $ represents one-way time-of-flight between the reader and the tag, $d$ the distance from the antenna to the tag, and $c$ the speed of light. ${\varphi _{Tx,Rx}}$ represents the phase shift introduced by the transceiver's hardware circuit and wired cables. $s_{ID}$ contains the tag's unique identification information using anti-collision UHF-RFID protocols \cite{Impinj2013}. Thus the equivalent baseband complex signal after coherent demodulation is given as
\begin{equation}\label{3}
s_{BS}(t)\!=\!{a_{BS}(t)}\exp\left\lbrace -j\!\left( {4\pi f\tau_0\!+\!{\varphi _{Tx,Rx}}\!+\!{\varphi _{Tag}}} \right)\right\rbrace,
\end{equation}where $a_{BS}$ is the amplitude, and ${\varphi _{Tag}}$ is the phase shift caused by tag's reflection characteristic and orientation.
\begin{figure}[t]
\centering
  \includegraphics[width=0.43\textwidth]{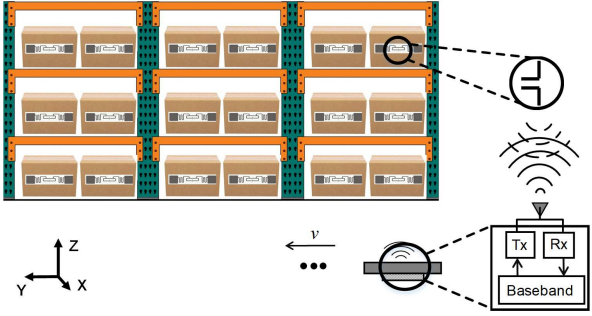}\\
  \caption{Conceptual diagram of setups and signal propagation between UHF-RFID reader and tags on the steel rack (Tx and Rx are reader transmitter and receiver, respectively.).  }
  \vspace{-0.4cm}
\end{figure}
\subsection{Analysis of the Measured Phase}
According to (3) considering no multipath interference, the phase extracted from the baseband signal can be given by
\begin{equation}\label{4}
\phi={4\pi\frac{d}{\lambda}\!+\!{\varphi _{Tx,Rx}}\!+\!{\varphi _{Tag}}},
\end{equation}where $\lambda=c/f$ is the wavelength. As we can see from (4), the phase is not only dependent on distance, but also the characteristic of the transceiver and RFID tag. Moreover, due to the modulo-$2\pi$ operation of RFID reader, the measured phases are wrapped within $[0,2\pi)$, namely, ${\phi _{m}} = \bmod \left( {\phi ,{\rm{ }}2\pi } \right)$. In this section, the basic investigation with regards to the tag's and antenna's orientation, and the distribution of measured phases will be presented.
\subsubsection{Orientation}
In the orientation experiment, three different RFID tags (namely, Alien G, SMARTRAC DogBone and SMARTRAC Belt), and the UHF RFID antenna Keonn Advantenna-SP11 are utilized. To investigate the characteristic of the orientation, the tag and antenna are mounted on the turntable. As the sketch map in Fig. 2 shows, the tags are rotated in three dimensions by 360 degrees, marked as X (roll), Y (pitch), and Z (yaw), while the antenna is rotated in two dimensions (Y and Z) by 180 degrees. In Fig. 2, the measured phase is more invariant when rotating the tag or antenna along Y/Z-axis, which has up to about 0.1 $\sim$ 0.15 rad fluctuations for a $\pm$45 degree rotation at 90 degrees. So the phase shift caused by the directivity misalignment of the antenna and the tag can be neglected. As for rotating the tag along X-axis, we obtain the $2\pi$ phase shift in case of 180-degrees rotation in Fig. 2(a), which means the phase uncertainty caused by the orientation can not be calibrated in advance.

\begin{figure}[t]
\centering
\subfigure[Tag's rotation]{
\includegraphics[width=0.22\textwidth]{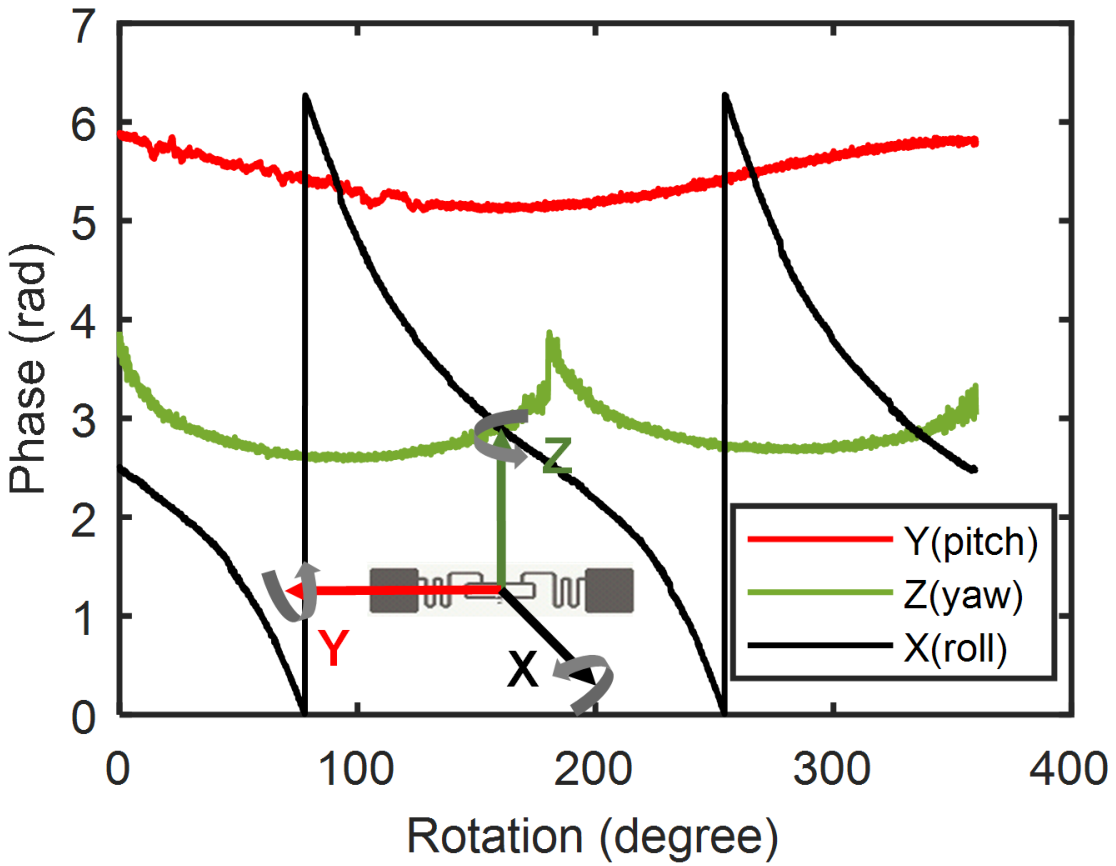}
}
\!
\subfigure[Antenna's rotation]{
\includegraphics[width=0.22\textwidth]{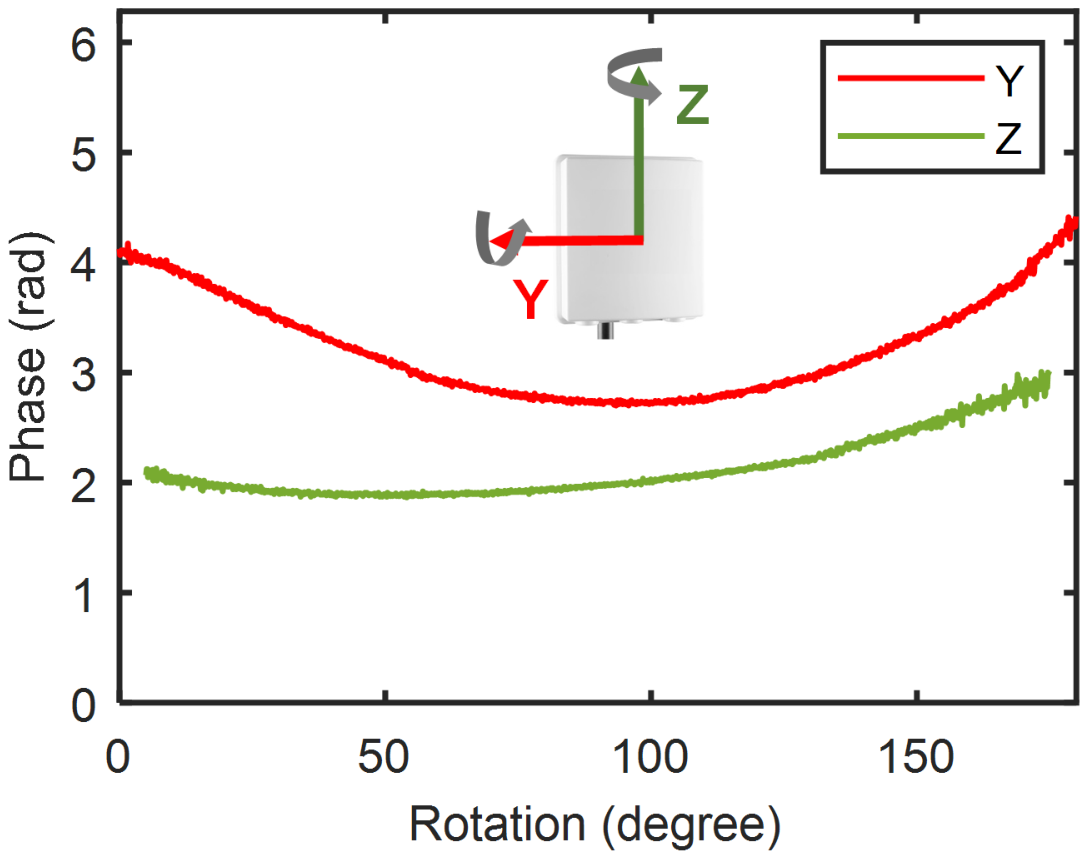}
}
\caption{Phase performance with the antenna's and tag's rotation.}
\vspace{-0.25cm}
\end{figure}

\begin{figure}[t]
\centering
\setlength{\abovecaptionskip}{-0.01cm}
\subfigure[]{
\includegraphics[width=0.23\textwidth]{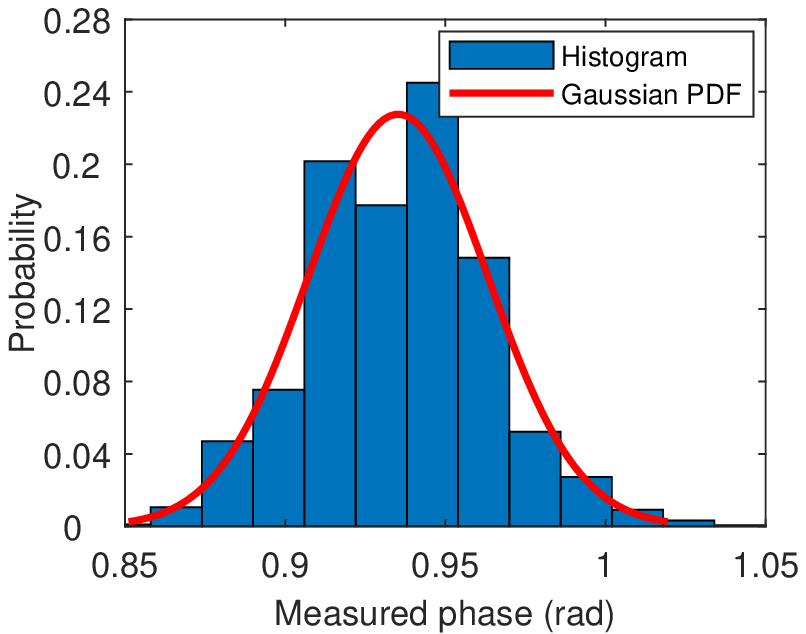}
}
\!\!\!\!\!\!
\subfigure[]{
\includegraphics[width=0.23\textwidth]{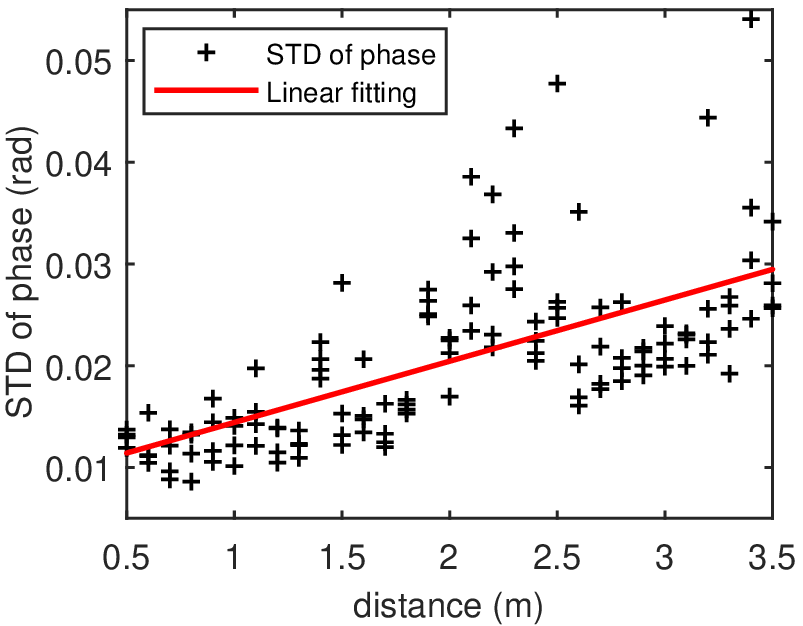}
}
\caption{Meaured phases: (a) the histogram of measured phase and Gaussian PDF fitting, (b) the STD of phase versus distance.}
\vspace{-0.5cm}
\end{figure}

\subsubsection{Distribution}
The measured phase is derived from the backscattered signal, which is always contaminated by the thermal noise and environmental clutter, leading to measurement errors. To investigate the statistical distribution of measured phase, we conduct the experiment with different types of tags, channels, and distances. In the experiment, the tags are also placed on the steel rack. Fig. 3(a) shows the normalized histogram of measured phases, as well as the Gaussian PDF fitting curve. The histogram has a good match with the Gaussian probability density function (PDF). So it is reasonable to assume that the measured phases follow a Gaussian distribution as $\mathcal{N}(\mu,\sigma^2)$, which also has been reported in \cite{Zhao2017,LiuTianCi,Tagoram}. Fig. 3(b) shows the standard deviation (STD) of the measured phases. It presents a slow increase with respect to the distance. The STD of the measured phases can be approximated by $\sigma=0.006d+0.0084$ rad using a linear fitting. Generally, the tag localization for inventory management is for short-range positioning with a range between 1.5 meters and 3 meters. Thus the STD has very slight variations by 0.01 rad. Therefore, for short-range UHF-RFID tag positioning, the STD of the measured phases can be regarded as a constant due to the small-amplitude fluctuations.

\section{Phase-based Positioning}
\subsection{Maximum Likelihood Estimation}
According to Section II, the measured phases of tag follows a Gaussian distribution, so we can use MLE to solve the positioning problem. Considering the $N$ independent observations of the measured phases, the MLE is given by
\begin{equation}\label{5}
{{\bf{P}}_{tag}}=\mathop {\arg \max }\limits_{{{\bf{P}}_{tag}}}\!\prod\limits_{n = 0}^{N\!-\!1}\!{\frac{1}{\sqrt{2\pi}\sigma[n]}e^{-\frac{{\left( {{\phi _{m}}\left[ n \right]-\!\bmod\!\left(\!{\frac{{4\pi }}{\lambda }{d}\left[n \right]+\!\varphi_0}\!\right)}\!\right)}^2}{2\sigma^2[n]}}\!}.
\end{equation}where $d=\Vert {{\bf{P}}_{ant}}-{{\bf{P}}_{tag}} \Vert$ is the distance from the antenna's coordinates ${{\bf{P}}_{ant}}$ to the tag's coordinates ${{\bf{P}}_{tag}}$. $\bmod(\cdot)$ is the modulo-$2\pi$ operator, and $\varphi_0=\!{\varphi _{Tx,Rx}}\!+\!{\varphi _{Tag}}$. With the assumption of short-range UHF-RFID as mentioned above, the STD can be regarded as a constant. So (5) can be rewritten as 
\begin{equation}\label{6}
\begin{aligned}
\!{{\bf{P}}_{tag}}\!&=\!\mathop {\arg \max }\limits_{{{\bf{P}}_{tag}}}\!\prod\limits_{n = 0}^{N\!-\!1}\!{\frac{1}{\sqrt{2\pi}\sigma}e^{-\frac{{\left( {{\phi _{m}}\left[ n \right]-\!\bmod\!\left(\!{\frac{{4\pi }}{\lambda }{d}\left[n \right]+\!\varphi_0}\!\right)}\!\right)}^2}{2\sigma^2}}\!}\\
\!&=\!\mathop {\arg \max }\limits_{{{\bf{P}}_{tag}}}\!\sum\limits_{n = 0}^{N\!-\!1}\!{\left[ {\!-\!{{\left( {{\phi_{m}}\!\left[ n \right]\!-\!\bmod\!\left(\!{\frac{{4\pi }}{\lambda }{d}\!\left[n \right]\!+\!\varphi_0}\!\right)}\!\right)}^2}}\!\right]}.
\end{aligned}
\end{equation}
\par
Due to the tag diversity, the orientation (X-axis rolling in Fig. 2(a)), and the frequency diversity \cite{Zhao2017}, the phase shift $\varphi_0$ cannot been calibrated before the positioning procedure. But for the same tag, $\varphi_o$ is almost constant when the antenna moves along the tag. It has less than 0.15-rad fluctuations as the result of directivity misalignment of antenna and tag, as shown in Fig. 2 (Y/Z-axis rotation). So in this section, a differential mitigation method is proposed to mitigate the uncertain phase shift $\varphi_o$. 
\par
The first scheme of the differential mitigation is the misaligned subtraction, in which we use the differences of the adjacent measured phases to eliminate the impact of $\varphi_0$. Define $\Delta {\phi _{m}^{[n,n\!-\!1]}}\!=\phi_{m}[n]\!-\phi_{m}[n-1]$, and $\!\Delta\varphi_d^{[n,n\!-\!1]}\!=\!\bmod\!\left(\!{\frac{{4\pi }}{\lambda }{d}\!\left[n \right]\!+\!\varphi_0}\!\right)\!\!-\!\!\bmod\!\left(\!{\frac{{4\pi}}{\lambda }{d}\!\left[n\!-\!1\right]\!+\!\varphi_0}\!\right)$, so we have
\begin{equation}\label{7}
{{\bf{P}}_{tag}}\!=\!\mathop {\arg \max }\limits_{{{\bf{P}}_{tag}}}\!\sum\limits_{n = 1}^{N\!-\!1}\!{\left[-\left({\Delta {\phi _{m}^{[n,n\!-\!1]}}\!-\!\Delta\varphi_d^{[n,n\!-\!1]}}\right)^2\right]}.
\end{equation}
If the spatial sampling interval satisfies $\frac{4\pi}{\lambda}\vert {\Delta d^{[n,n\!-\!1]}}\vert<2\pi$, where ${\Delta d^{[n,n\!-\!1]}}\!=d[n]-\!d[n-1]$, then $\Delta\varphi_d^{[n,n\!-\!1]}$ can be given by \cite{LiuTianCi,UnwrappingRFIDTA}
\begin{equation}\label{8}
\!\Delta\varphi_d^{[n,n\!-\!1]}\!\!=\!\!\begin{cases} 
       \!\!\frac{{4\pi}}{\lambda}\Delta d^{[n,n\!-\!1]},\!\!
       			&\!\!\Delta d^{[n,n\!-\!1]}\cdot\Delta{\phi _{m}^{[n,n\!-\!1]}}>0 \\
      \!\!\frac{{4\pi}}{\lambda}\Delta d^{[n,n\!-\!1]}\!+\!2\pi,\!\!
      			&\!\!\Delta d^{[n,n\!-\!1]}\!<\!0,\Delta{\phi _{m}^{[n,n\!-\!1]}}\!>\!0\\
      \!\!\frac{{4\pi}}{\lambda}\Delta d^{[n,n\!-\!1]}\!-\!2\pi,\!\!
      			&\!\!\Delta d^{[n,n\!-\!1]}\!>\!0,\Delta{\phi _{m}^{[n,n\!-\!1]}}\!<\!0\\
   \end{cases}\!.
\end{equation}
However, the above spatial sampling interval is not always satisfied considering the fast moving RFID antenna, channel fading, read mode of RFID reader, etc. So we make a modulo operation to eliminate the constraint, given as
\begin{equation}\label{9}
\Delta\varphi_d^{[n,n\!-\!1]}\!=\!\!\begin{cases} 
       \bmod\!\left(\!\frac{{4\pi }}{\lambda }\Delta d^{[n,n-\!1]}\right)-\!2\pi,\!\!
       & \Delta\varphi_d^{[n\!-\!1]}\!\in\!(-2\pi,0) \\
      \bmod\!\left(\!\frac{{4\pi }}{\lambda }\Delta d^{[n,n-\!1]}\right),\!\! 
      & \Delta\varphi_d^{[n\!-\!1]}\!\in\![0,2\pi)
   \end{cases}\!.
\end{equation}
\par
The second scheme is selecting one of the measured phases as the reference $\phi_{m}[r]$, such as the phase at the first recorded position. Thus the tag's position can be given as
\begin{equation}\label{10}
{{\bf{P}}_{tag}}\!=\!\mathop {\arg \max }\limits_{{{\bf{P}}_{tag}}}\!\sum\limits_{n = 0}^{N\!-\!1}\!{\left[-\left({\Delta {\phi _{m}^{[n,r]}}\!-\!\Delta\varphi_d^{[n,r]}}\right)^2\right]},
\end{equation} where $\Delta {\phi _{m}^{[n,r]}}\!=\phi_{m}[n]\!-\phi_{m}[r]$, and 
\begin{equation}\label{11}
\!\Delta\varphi_d^{[n,r]}\!=\!\!\begin{cases} 
       \bmod\!\left(\!\frac{{4\pi }}{\lambda }\Delta d^{[n,r]}\right)\!-\!2\pi,\!\!
       & \Delta\varphi_d^{[n,r]}\!\in\!(-2\pi,0) \\
      \bmod\!\left(\!\frac{{4\pi }}{\lambda }\Delta d^{[n,r]}\right),\!\! 
      & \Delta\varphi_d^{[n,r]}\!\in\![0,2\pi)
   \end{cases}\!.
\end{equation}With the conversion in (9) and (11), the phase uncertainty is mitigated. However, the judging condition in (9) and (11), namely $\Delta\varphi_d^{[n,n\!-\!1]}\gtrless 0$ or $\!\Delta\varphi_d^{[n,r]}\gtrless 0$, is an unknown prophet.

\begin{figure}[t]
\centering
 \vspace{0.1cm}
 \setlength{\abovecaptionskip}{-0.1cm}
  \includegraphics[width=0.36\textwidth]{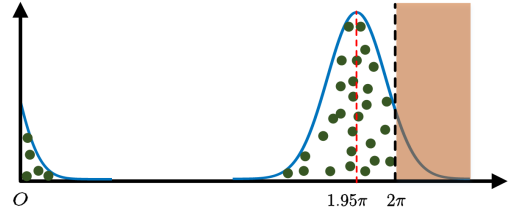}\\
  \caption{An example of phase jump when the actual phase is $1.95\pi$ rad.
  }
 \vspace{-0.45cm}
\end{figure}

\subsection{Likelihood Function Reconstruction}
Low level user data, phase captured by commercial off-the-shelf (COTS) UHF-RFID reader, brings opportunities to realize fine-grained localization. But it also brings a tricky problem, phase jumps, as a result of the modulo-$2\pi$ operation. As shown in Fig. 4, when the true phase is very close to $2\pi$ rad (or $0$ rad), the measured phase may jump to the value left to the $0$ rad (or right to the $2\pi$ rad).
So the likelihood function $f_{NLF}(\Delta\phi _{m}^{[n,r]}|d)=-\left({\Delta {\phi _{m}^{[n,r]}}\!-\!\Delta\varphi_d^{[n,r]}}\right)^2$ in (10)\footnote{Notice that here and in the following, we only present the likelihood function in case of reference subtraction, while it has the same form for misalignment subtraction, namely, $f(\Delta\phi _{m}^{[n,n-\!1]}|d)$.}, say the naive likelihood function (NLF), will cause large errors as a result of $\Delta {\phi _{m}^{[n,r]}}$ abrupt jumping when the measured phase is at around $2\pi$ or $0$ rad. For example, when the actual measured phase $\phi _{m}[r]=1.6\pi$ rad and $\phi _{m}[n]=1.95\pi$ rad, the phase difference is $\Delta {\phi _{m}^{[n,r]}}=0.35\pi$ rad. But due to noise or other interference, $\phi _{m}[n]$ may jump to $0.03\pi$ rad, and then $\Delta {\phi _{m}^{[n,r]}}=-1.57\pi$ rad, which brings a large offset to the likelihood function (NLF). 
\par
To cope with the discontinuities caused by the phase jump, a trigonometric function transformation is introduced. The \textit{cosine} function is a good choice, and makes the function values before and after $2\pi$ or $0$ rad approaching to each other. Meanwhile, it is interesting to find that NLF in (10) has a good match with \textit{cosine} function when utilizing the second-order Taylor series approaching method, so the maximum likelihood positioning estimation can be converted as
\begin{equation}\label{12}
\begin{aligned}
{{\bf{P}}_{tag}}\!&=\!\mathop {\arg \max }\limits_{{{\bf{P}}_{tag}}}\!\sum\limits_{n = 0}^{N\!-\!1}\!{\left[-\left({\Delta {\phi _{m}^{[n,r]}}\!-\!\Delta\varphi_d^{[n,r]}}\right)^2\right]}\\
&\doteq\mathop {\arg \max }\limits_{{{\bf{P}}_{tag}}} \sum\limits_{n = 0}^{N - 1} {\underbrace {{\cos \left({\Delta {\phi _{m}^{[n,r]}}\!-\!\Delta\varphi_d^{[n,r]}}\right)}}_{f_{CLF}(\Delta\phi _{m}^{[n,r]}|d)}}\!,\!
\end{aligned}
\end{equation}where the reconstructed likelihood function in (12) is defined as \textit{cosine} likelihood function (CLF), marked as $f_{CLF}(\Delta\phi _{m}|d)$. Moreover, it should be noted that the \textit{cosine} transformation also mitigates the condition judgment (the prophet) in (11), since the $-2\pi$ compensation when $\Delta\varphi_d^{[n,r]}\!\in\!(-2\pi,0)$ will not change the value of the likelihood function as a result of \textit{cosine} transformation, namely, $\cos \left({\Delta {\phi _{m}^{[n,r]}}\!-\!\Delta\varphi_d^{[n,r]}}\right)=\cos \left({\Delta {\phi _{m}^{[n,r]}}\!-\!\frac{{4\pi }}{\lambda }\Delta d^{[n,r]}}\right)$. Likewise, we can also operate a \textit{sine} transformation with respect to NLF with the Taylor series approaching, namely \textit{sine} likelihood function (SLF), defined by $f_{SLF}(\Delta\phi _{m}|d)$. So the positioning estimation is given as
\begin{equation}\label{13}
\begin{aligned}
\!\!{{\bf{P}}_{tag}}\!&=\!\mathop {\arg \max }\limits_{{{\bf{P}}_{tag}}}\!\sum\limits_{n = 0}^{N\!-\!1}\!{\left[-\left({\Delta {\phi _{m}^{[n,r]}}\!-\!\Delta\varphi_d^{[n,r]}}\right)^2\right]}\\
\!&\doteq\mathop {\arg \max }\limits_{{{\bf{P}}_{tag}}} \sum\limits_{n = 0}^{N\!-\!1} {\underbrace {{\left[-\!\sin\left({\!\Delta {\phi _{m}^{[n,r]}}\!-\!\frac{{4\pi }}{\lambda }\Delta d^{[n,r]}}\right)^2\right]}}_{f_{SLF}(\Delta\phi _{m}^{[n,r]}|d)}}\!.\!
\end{aligned}
\end{equation}
\par
As for practical applications in warehouses, the UHF-RFID signal may contain contaminated components from the thermal noise, reflection, scattering or other interference. So the contaminated phase will contain an offset with the actual phase. To cover this, a weighted MLE is proposed to augment the likelihood function. The component with a smaller bias compared with the calculated phase will be assigned a larger weight. The weighted MLE can be given as
\begin{equation}\label{14}
{{\bf{P}}_{tag}}\!=\!\mathop {\arg \max }\limits_{{{\bf{P}}_{tag}}}\!\sum\limits_{n = 0}^{N\!-\!1}\!w_i[n]f_{i}(\Delta\phi _{m}^{[n,r]}|d),
\end{equation}where $i=\{CLF, SLF\}$. $w_i[n]\in[0,1]$ is constructed based on the idea that the measured phase with a larger offsets of $\Delta\phi _{m}^{[n,r]}$ towards $\Delta\varphi_d^{[n,r]}$ will be assigned a smaller weight. Thus the weights can be defined by
\begin{equation}\label{15}
w_i[n]\!=\!\!\begin{cases} 
       \vert f_{CLF}(\Delta\phi _{m}^{[n,r]}|d)\vert,\!\!
       & i=CLF \\
      e^{f_{SLF}(\Delta\phi _{m}^{[n,r]}|d)},\!\! & i=SLF
   \end{cases}\!.
\end{equation}

\section{Microbenchmark}
\subsection{Configuration}
To investigate the localization accuracy of the proposed method, a phase-based UHF-RFID positioning system is established. In the positioning system, the Impinj Speedway R420 RFID reader is adopted without any hardware modification. The reader connect with the PC controller through the Ethernet cable under the LLRP protocol \cite{Impinj2013}. The antenna involved is Keonn Advantenna-SP11, which is circular polarization with the gain 8.3 dBi. Antenna SP11 is installed on the linear track of the Velmex system, which is driven by the PC controller via MatLab. 14 Alien G tags are attached to the boxes on the steel rack, as shown in Fig. 5. Considering single one RFID antenna is adopted in the experiment, the tags on the first two levels are considered to avoid position ambiguity along the altitude\footnote{When the antenna moves along a linear trajectory, it will produce a symmetric position ambiguity with respect to the trajectory, which has the same distance to the antenna as the actual position.}. The distance from the antenna to the rack is 1.4 meters. The frequency is 866.9 MHz. Define that X-axis is vertical to the plane of the rack, Y-axis and Z-axis are along the linear track and the altitude, respectively. 

\begin{figure}[t]
\centering
\setlength{\abovecaptionskip}{-0.2cm}
  \includegraphics[width=0.32\textwidth]{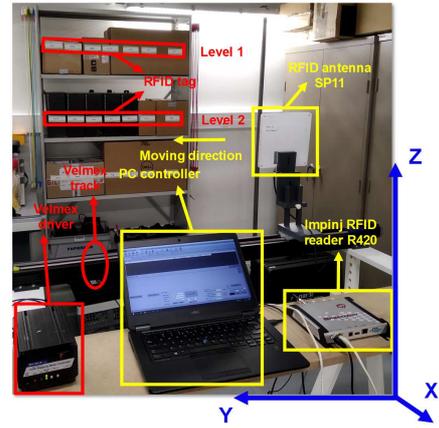}\\
  \caption{The setups of UHF-RFID positioning system.}
  \vspace{-0.5cm}
\end{figure}

\subsection{Performance Evaluation}
\begin{figure*}[t]
\centering
\setlength{\abovecaptionskip}{-0.03cm}
\setlength{\belowcaptionskip}{-0.05cm}
\subfigure[CLF misaligned subtraction]{
\includegraphics[width=0.232\textwidth]{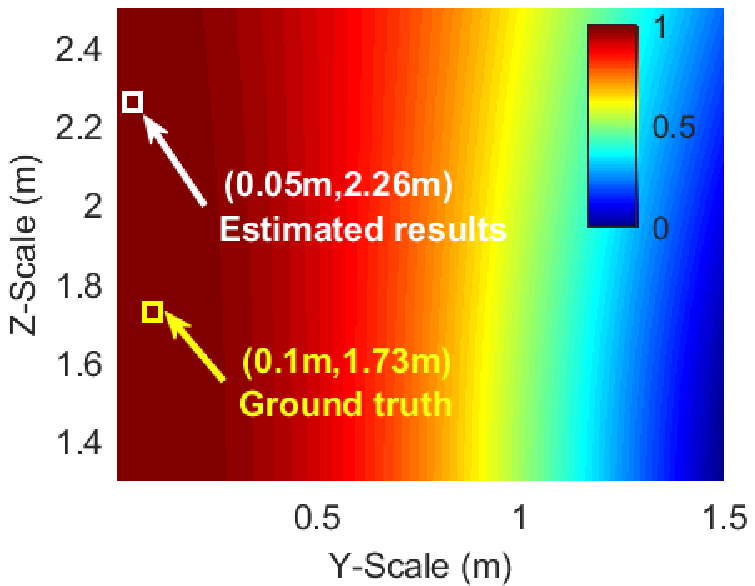}
}
\!\!\!\!\!\!
\subfigure[SLF misaligned subtraction]{
\includegraphics[width=0.232\textwidth]{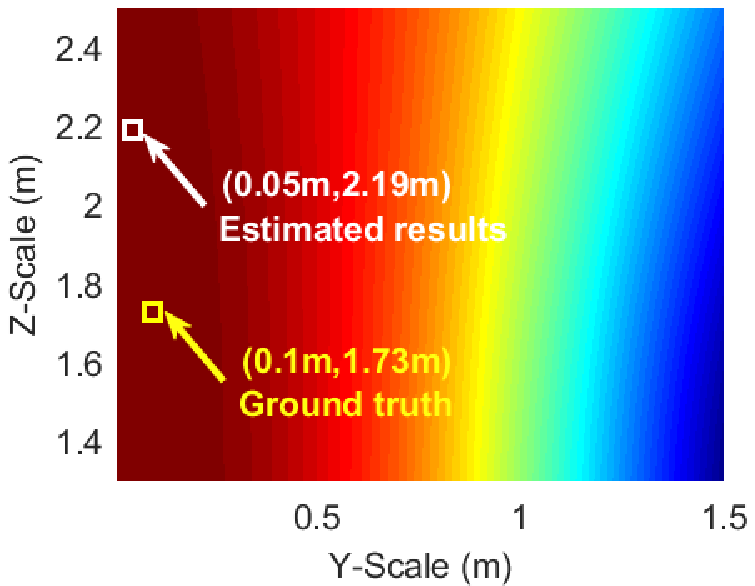}
}
\!\!\!\!\!\!
\subfigure[CLF reference subtraction]{
\includegraphics[width=0.232\textwidth]{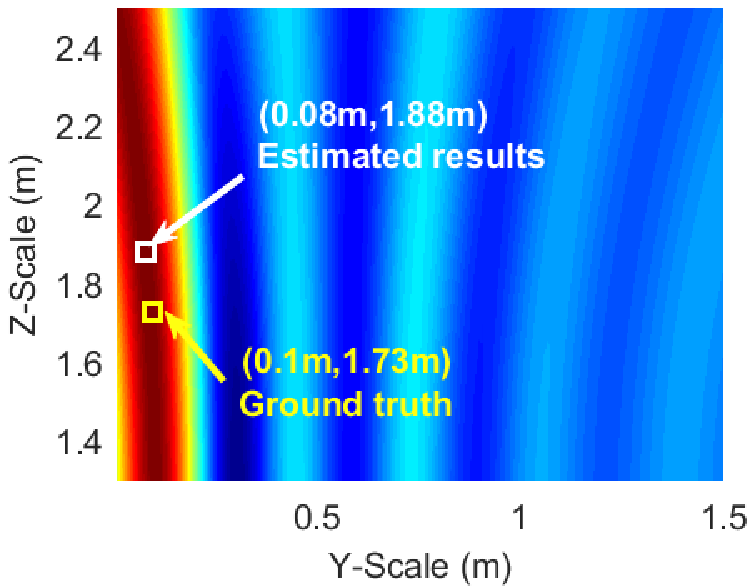}
}
\!\!\!\!\!\!
\subfigure[SLF reference subtraction]{
\includegraphics[width=0.232\textwidth]{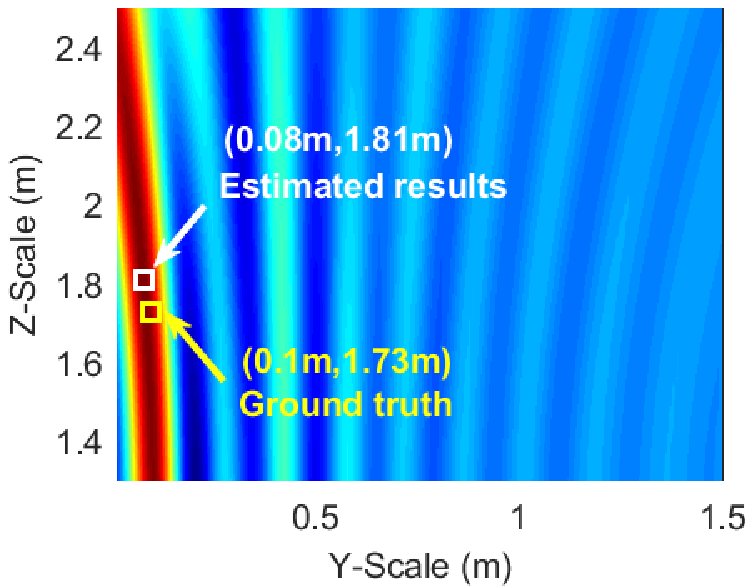}
}
\!\!
\subfigure[Weighted CLF]{
\includegraphics[width=0.232\textwidth]{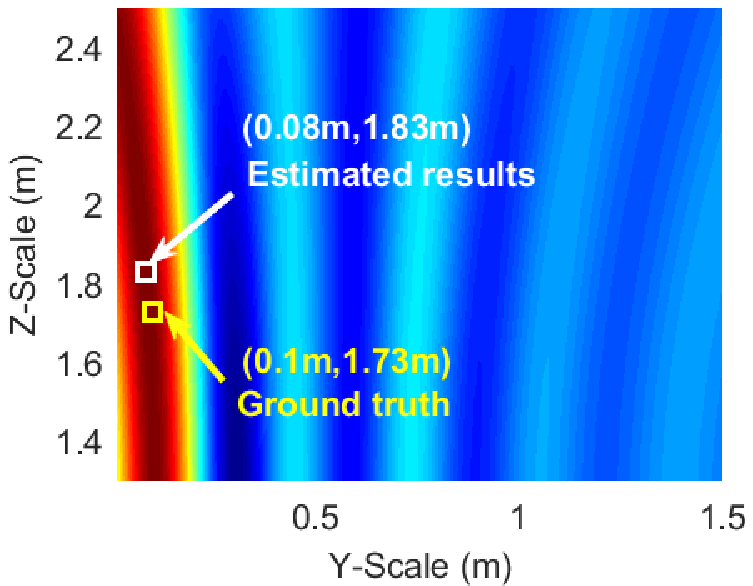}
}
\!\!\!\!\!\!
\subfigure[Weighted SLF]{
\includegraphics[width=0.232\textwidth]{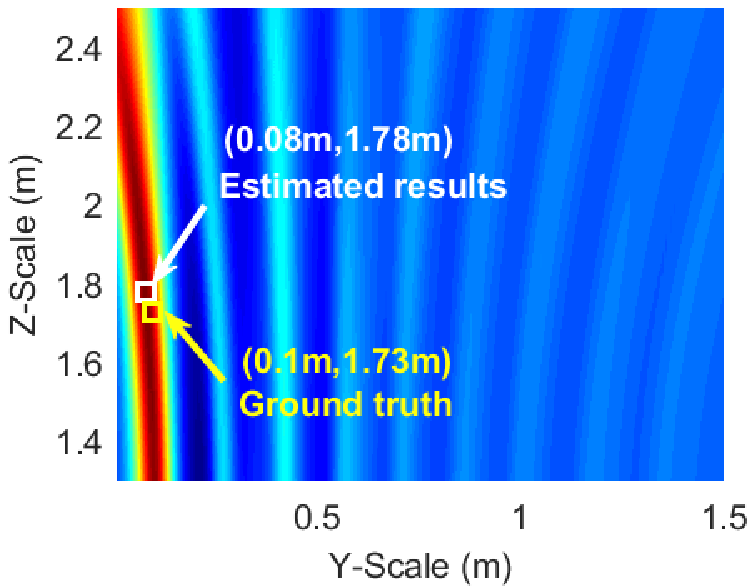}
}
\!\!\!\!\!\!
\subfigure[SARFID]{
\includegraphics[width=0.232\textwidth]{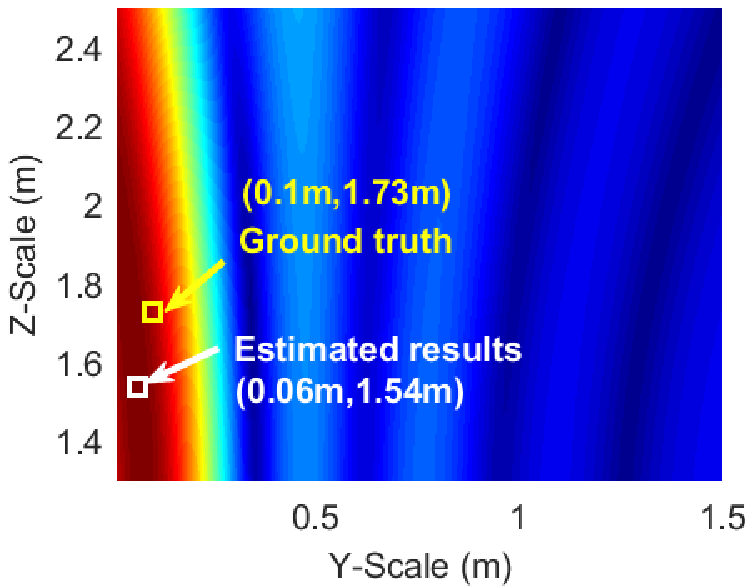}
}
\!\!\!\!\!\!
\subfigure[Tagoram]{
\includegraphics[width=0.232\textwidth]{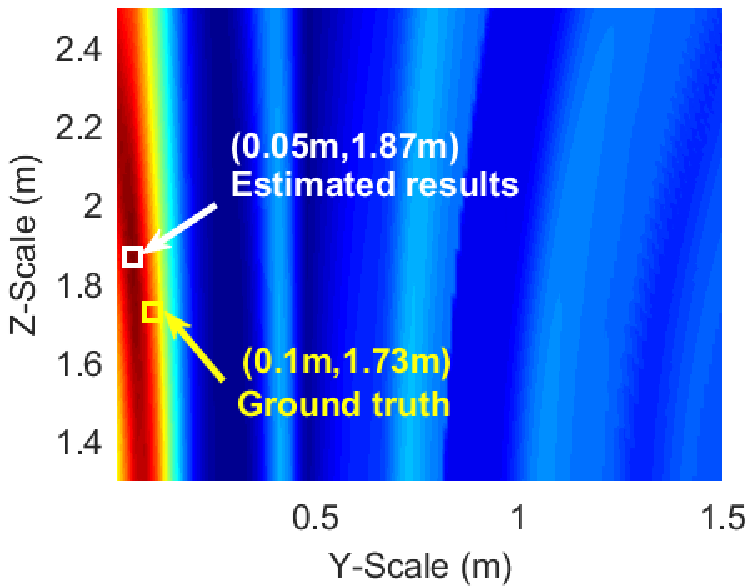}
}
\!\!
\caption{The normalized hologram at the plane of the steel rack under different methods (The first tag on the level one in Fig. 5 is selected).}
\vspace{-0.3cm}
\end{figure*}

\begin{figure}[t]
\centering
\vspace{-0.1cm}
\setlength{\abovecaptionskip}{-0.05cm}
  \includegraphics[width=0.45\textwidth]{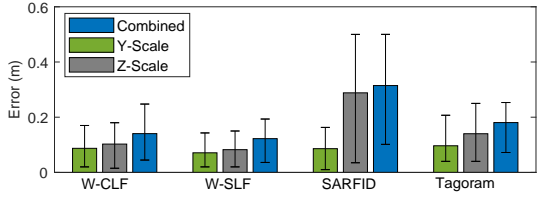}\\
  \caption{Accuracy comparison with two state-of-art methods (\textit{combined} represents the combination of Y-scale and Z-scale errors).}
  \vspace{-0.4cm}
\end{figure}

To mitigate the impact caused by phase shift $\varphi_0$, two differential augmentation schemes have been proposed: misaligned subtraction, and reference subtraction (Section III-A). As for the reference subtraction, the measured phase at the first position is chosen as the reference in this experiment. As shown in Fig. 6(a)-(d), we compare the positioning performance of the two schemes based on the normalized hologram at the plane of the steel rack (namely, Y-scale and Z-scale). It is obvious that the reference subtraction scheme (about 15.1 $cm$ and 8.2 $cm$ positioning errors for CLF- and SLF-based method, respectively) can obtain a distinct accuracy improvement compared with the misaligned subtraction (about 53.2 $cm$ and 46.3 $cm$ errors, respectively). Moreover, reference subtraction scheme mitigates most of the deceptive positions with high likelihood, namely narrower candidate regions. We also observe that the SLF-based reference subtraction outperforms CLF-based method slightly with accuracy increasing by about 7 $cm$.
\par
Considering that the measured phases may suffer from the unexpected interference, the weights are constructed to augment MLE positioning performance in (15). To evaluate the positioning accuracy further, we also compare the proposed methods with two state-of-art methods: SARFID \cite{Buffi2015,Buffi2017} and Tagoram\cite{Tagoram}. As shown in Fig. 6(e)-(h), our methods realize the localization with small errors, especially for weighted SLF (about 5.4 $cm$ positioning offsets towards the ground truth), while SARFID and Tagoram have 19.4 $cm$ and 14.9 $cm$ offsets, respectively. Fig. 7 presents the statistical errors of weighted CLF (W-CLF), weighted SLF (W-SLF), SARFID, and Tagoram. We can see that the weighted methods (W-CLF, W-SLF, and Tagoram\footnote{Tagoram is weighted by the cumulative distribution function (CDF) of the measured phases in \cite{Tagoram}.}) achieve higher accuracy and have better tolerance towards the unexpected interference. The proposed W-CLF and W-SLF realize fine-grained positioning results with the mean errors of 14.2 $cm$ and 11.6 $cm$. Meanwhile, W-SLF is the best performer in this case, which achieves 33\% and 58\% accuracy improvement compared with Tagoram and SARFID, respectively. Moreover, it should be noted that we only use one antenna in the experiment. More antennas placed along the altitude (or Z-scale motion) will improve the positioning performance in the Z-scale as a result of the increasing size of virtual synthetic aperture.
\section{Conclusion}
In this paper, we propose a phase-based UHF-RFID tag positioning method based on weighted variant MLE. To mitigate the intrinsic phase uncertainty, the likelihood function has been reconstructed based on trigonometric convection. Then the weights have been constructed to improve the positioning accuracy. The positioning performance has been evaluated through experiment with a single RFID antenna. The results indicate that our proposed methods can realize high localization accuracy with the mean errors of 14.2 $cm$ and 11.6 $cm$ for W-CLF and W-SLF, respectively. Especially for W-SLF, it achieves 33\% and 58\% accuracy improvement compared with the existing methods, such as Tagoram and SARFID, respectively.
\section*{Acknowledgment}
This work is supported in part by the Excellence of Science (EOS) project MUlti-SErvice WIreless NETworks (MUSE-WINET), and IMEC co-financed project InWareDrones.

\bibliographystyle{IEEEtran}
\bibliography{PhaseBasedLoc}

\begin{thebibliography}{10}
\providecommand{\url}[1]{#1}
\csname url@samestyle\endcsname
\providecommand{\newblock}{\relax}
\providecommand{\bibinfo}[2]{#2}
\providecommand{\BIBentrySTDinterwordspacing}{\spaceskip=0pt\relax}
\providecommand{\BIBentryALTinterwordstretchfactor}{4}
\providecommand{\BIBentryALTinterwordspacing}{\spaceskip=\fontdimen2\font plus
\BIBentryALTinterwordstretchfactor\fontdimen3\font minus
  \fontdimen4\font\relax}
\providecommand{\BIBforeignlanguage}[2]{{%
\expandafter\ifx\csname l@#1\endcsname\relax
\typeout{** WARNING: IEEEtran.bst: No hyphenation pattern has been}%
\typeout{** loaded for the language `#1'. Using the pattern for}%
\typeout{** the default language instead.}%
\else
\language=\csname l@#1\endcsname
\fi
#2}}
\providecommand{\BIBdecl}{\relax}
\BIBdecl

\bibitem{Zhang2014}
Z.~{Zhang}, Z.~{Lu}, V.~{Saakian}, X.~{Qin}, Q.~{Chen}, and L.~{Zheng},
  ``Item-level indoor localization with passive uhf rfid based on tag
  interaction analysis,'' \emph{IEEE Transactions on Industrial Electronics},
  vol.~61, no.~4, pp. 2122--2135, Apr 2014.

\bibitem{Khan2009}
M.~A. {Khan} and V.~K. {Antiwal}, ``Location estimation technique using
  extended {3-D LANDMARC} algorithm for passive rfid tag,'' in \emph{IEEE
  International Advance Computing Conference}, Mar 2009, pp. 249--253.

\bibitem{Nikitin2010}
P.~V. Nikitin, R.~Martinez, S.~Ramamurthy, H.~Leland, G.~Spiess, and K.~V. Rao,
  ``{Phase based spatial identification of UHF RFID tags},'' in \emph{IEEE
  International Conference on RFID}, May 2010, pp. 102--109.

\bibitem{Buffi2015}
A.~Buffi, P.~Nepa, and F.~Lombardini, ``{A phase-based technique for
  localization of UHF-RFID tags moving on a conveyor belt: Performance analysis
  and test-case measurements},'' \emph{IEEE Sensors Journal}, vol.~15, no.~1,
  pp. 387--396, Jan 2015.

\bibitem{Buffi2017}
A.~{Buffi} and P.~{Nepa}, ``The sarfid technique for discriminating tagged
  items moving through a uhf-rfid gate,'' \emph{IEEE Sensors Journal}, vol.~17,
  no.~9, pp. 2863--2870, May 2017.

\bibitem{LiuTianCi}
T.~{Liu}, L.~{Yang}, Q.~{Lin}, Y.~{Guo}, and Y.~{Liu}, ``Anchor-free
  backscatter positioning for rfid tags with high accuracy,'' in \emph{IEEE
  Conference on Computer Communications}, Apr 2014, pp. 379--387.

\bibitem{Tanghe2008}
E.~{Tanghe}, W.~{Joseph}, L.~{Verloock}, L.~{Martens}, H.~{Capoen}, K.~V.
  {Herwegen}, and W.~{Vantomme}, ``The industrial indoor channel: large-scale
  and temporal fading at 900, 2400, and 5200 mhz,'' \emph{IEEE Transactions on
  Wireless Communications}, vol.~7, no.~7, pp. 2740--2751, Jul 2008.

\bibitem{Zhao2017}
R.~Zhao, Q.~Zhang, D.~Li, H.~Chen, and D.~Wang, ``{A novel accurate synthetic
  aperture RFID localization method with high radial accuracy},'' \emph{18th
  IEEE International Symposium on A World of Wireless, Mobile and Multimedia
  Networks, WoWMoM 2017}, pp. 1--9, Jul 2017.

\bibitem{Tagoram}
L.~Yang, Y.~Chen, X.-Y. Li, C.~Xiao, M.~Li, and Y.~Liu, ``Tagoram: Real-time
  tracking of mobile rfid tags to high precision using cots devices,'' in
  \emph{Proceedings of the 20th Annual International Conference on Mobile
  Computing and Networking}, ser. MobiCom '14, Sep 2014, pp. 237--248.

\bibitem{Shangguan2016}
L.~Shangguan and K.~Jamieson, ``The design and implementation of a mobile rfid
  tag sorting robot,'' in \emph{Proceedings of the 14th Annual International
  Conference on Mobile Systems, Applications, and Services}, ser. MobiSys '16,
  Jun 2016, pp. 31--42.

\bibitem{Impinj2013}
Impinj, ``Speedway revolution reader application note - low level user data
  support (revision 3.0, 2013),'' [Online]. Available:
  \url{https://support.impinj.com/hc/en-us/article
  attachments/200774268/SR_AN_IPJ_Speedway_Rev_Low_Level_Data_Support_20130911.pdf}.

\bibitem{UnwrappingRFIDTA}
C.~{Li}, E.~{Tanghe}, D.~{Plets}, P.~{Suanet}, J.~{Hoebeke}, E.~{De Poorter},
  and W.~{Joseph}, ``{RePos}: Relative position estimation of uhf-rfid tags for
  item-level localization,'' in \emph{IEEE International Conference on RFID
  Technology and Applications (RFID-TA)}, Sep 2019, pp. 357--361.

\end{thebibliography}

\end{document}